\documentclass[aps,twocolumn,superscriptaddress,prapplied,longbibliography]{revtex4-2}
\usepackage{amsmath}
\usepackage{amssymb}
\usepackage{amsfonts}
\usepackage{graphicx}
\usepackage{float}
\usepackage[margin=0.75in]{geometry}

\renewcommand{\bm}{\mathbf{m}}
\newcommand{\btau}{\boldsymbol{\tau}}
\newcommand{\xiFMR}{\xi_\text{FMR}}
\newcommand{\xiDL}{\xi_\text{DL}}
\renewcommand{\Re}[1]{\operatorname{Re}\left[ #1 \right]}

\begin{document}

\title{Resonant Measurement of Non-Reorientable Spin-Orbit Torque from a Ferromagnetic Source Layer  Accounting for Dynamic Spin Pumping}
\author{Joseph A. Mittelstaedt}
    \affiliation{Cornell University, Ithaca, NY 14850, USA}
\author{Daniel C. Ralph}
    \affiliation{Cornell University, Ithaca, NY 14850, USA}
    \affiliation{Kavli Institute at Cornell, Ithaca, NY 14853, USA}

\date{\today} 

\begin{abstract}
Using a multilayer structure containing (cobalt detector layer)/(copper spacer)/(Permalloy source layer), we show experimentally how the non-reorientable spin-orbit torque generated by the Permalloy source layer (the component of spin-orbit torque that does not change when the Permalloy magnetization is rotated) can be measured using spin-torque ferromagnetic resonance (ST-FMR) with lineshape analysis.  We find that dynamic spin pumping between the magnetic layers exerts torques on the magnetic layers as large or larger than the spin-orbit torques, so that if dynamic spin pumping is neglected the result would be a large overestimate of the spin-orbit torque. Nevertheless, the two effects can be separated by performing ST-FMR as a function of frequency.  We measure a non-reorientable spin torque ratio $\xi_\text{Py} = 0.04 \pm 0.01$ for spin current flow from Permalloy through an 8 nm Cu spacer to the Co, and a strength of dynamic spin pumping that is consistent with previous measurements by conventional ferromagnetic resonance. 
\end{abstract}

\maketitle

\section{Introduction}

Current-induced spin-orbit torques offer the potential for efficient manipulation of nanoscale magnets for memory applications \cite{Brataas2012, Wang2013, Oboril2015}.  
Among the families of materials that are of interest as potential sources of spin-orbit torque are metallic ferromagnets \cite{Taniguchi2015, Humphries2017, Amin2018, Iihama2018, Baek2018, Gibbons2018a, Das2018, Safranski2018, Ou2019, Keller2019, Amin2019a, Seki2019a, Ma2020, Safranski2020, Hibino2020, Koike2020}. 
They are predicted to generate current-induced spin currents with both a reorientable component in which the spin is always parallel to the source-layer magnetization, plus a non-reorientable component (with a fixed spin direction in-plane and perpendicular to the charge current) that, remarkably, is not dephased within the source magnet even though the spin is not aligned with the magnetization \cite{Amin2019a}.
The non-reorientable component has the same symmetries as the torque generated by the spin Hall effect from a non-magnetic source material. 

Previously, spin-orbit torques from metallic ferromagnets have been probed indirectly by exciting nonlocal spin currents through magnetic insulators \cite{Das2017, Das2018, Cramer2019, Omori2019, Wimmer2019}, and the reciprocal process of spin-to-charge conversion has been measured for spin currents injected from an insulating ferromagnetic into a metallic ferromagnet \cite{Miao2013, Wang2014, Du2014, Wu2015, Wahler2016, Tian2016}.  
However, it is more challenging to make direct measurements of spin-orbit torques generated by ferromagnets in magnetic trilayer structures that would be necessary for memory applications (i.e., ferromagnetic spin-source layer / non-magnetic spacer / ferromagnetic free layer to be manipulated).  
Ideally, one would like to apply a current and make quantitative measurements of just the deflection of the free-layer magnetization.  
However, in structures with two magnetic layers an applied current will exert torques on both layers and their dynamics will also be coupled by interlayer interactions, making it challenging to isolate just the strength of the spin-orbit torque acting on the free layer.  
For the reorientable component of the spin torque, one approach to address this challenge is to perform spin-torque ferromagnetic resonance (ST-FMR) experiments and analyze the dependence on dc current of the separate linewidths for each magnetic layer \cite{Iihama2018, Safranski2018,Seki2019a, Safranski2020, Hibino2020, Koike2020}, although care should be taken in avoiding artifacts that can affect this method \cite{Xu2020, Karimeddiny2021}.  
Here, we demonstrate that ST-FMR can also be used to achieve quantitative measurements of the non-reorientable (independent of the source-layer magnetization) component of spin-orbit torque generated by a ferromagnetic metal layer in a magnetic trilayer, by analyzing the resonant amplitudes and lineshapes. 
For a correct analysis of trilayers containing two metallic magnets (e.g., Co and Permalloy (Py)), it is critical to take into account that dynamic spin pumping \cite{Heinrich2003, Pogoryelov2020} couples the two magnetic layers and alters their resonance amplitudes.  
Still, the effects of spin currents associated with spin-orbit torques and those arising from dynamic spin pumping can be separated by analyzing the frequency dependence of the ST-FMR signals. 
Therefore, ST-FMR performed on magnetic trilayers can provide quantitative measurements of both the non-reorientable spin-orbit torque produced by a magnetic source layer and dynamic spin pumping.  Our approach is similar to that of Yang et al.\ \cite{Yang2020}, except that their method does not account for dynamic spin pumping.

\section{Sample Fabrication and Measurements}\label{sec:measurements}

We used DC magnetron sputtering onto thermally oxidized high-resistivity silicon wafers to grow heterostructures of containing both a Co ``detector" layer, whose resonance dynamics we analyze, and  a Ni$_{81}$Fe$_{19}$ (Py)  ``source" layer that generates a spin current to act on the Co.  Specifically, the sample layer structure is
 SiO$_x$/Ta (1)/Cu (8)/Co (8)/Cu (8)/Py ($t_\text{Py})$/Ta (1) with numbers in parentheses indicating thickness in nm.  The two Cu spacer layers have the same thickness, and are designed so that the in-plane component of the Oersted field that they generate will cancel within the Co detector, and consequently the only net in-plane Oersted field acting on the Co layer will be due to current flowing within the Py layer.  This will simplify our analysis of the spin-orbit torque acting on the Co layer. 
Cu was chosen for the spacer layers because of its long spin diffusion length and negligible spin Hall effect \cite{Bass2007}.  The Ta layers provide layer-smoothing and protection against oxidation, and have sufficiently large resistivity that they carry negligible current.

\begin{figure*}
\begin{center}
\includegraphics[width=\linewidth]{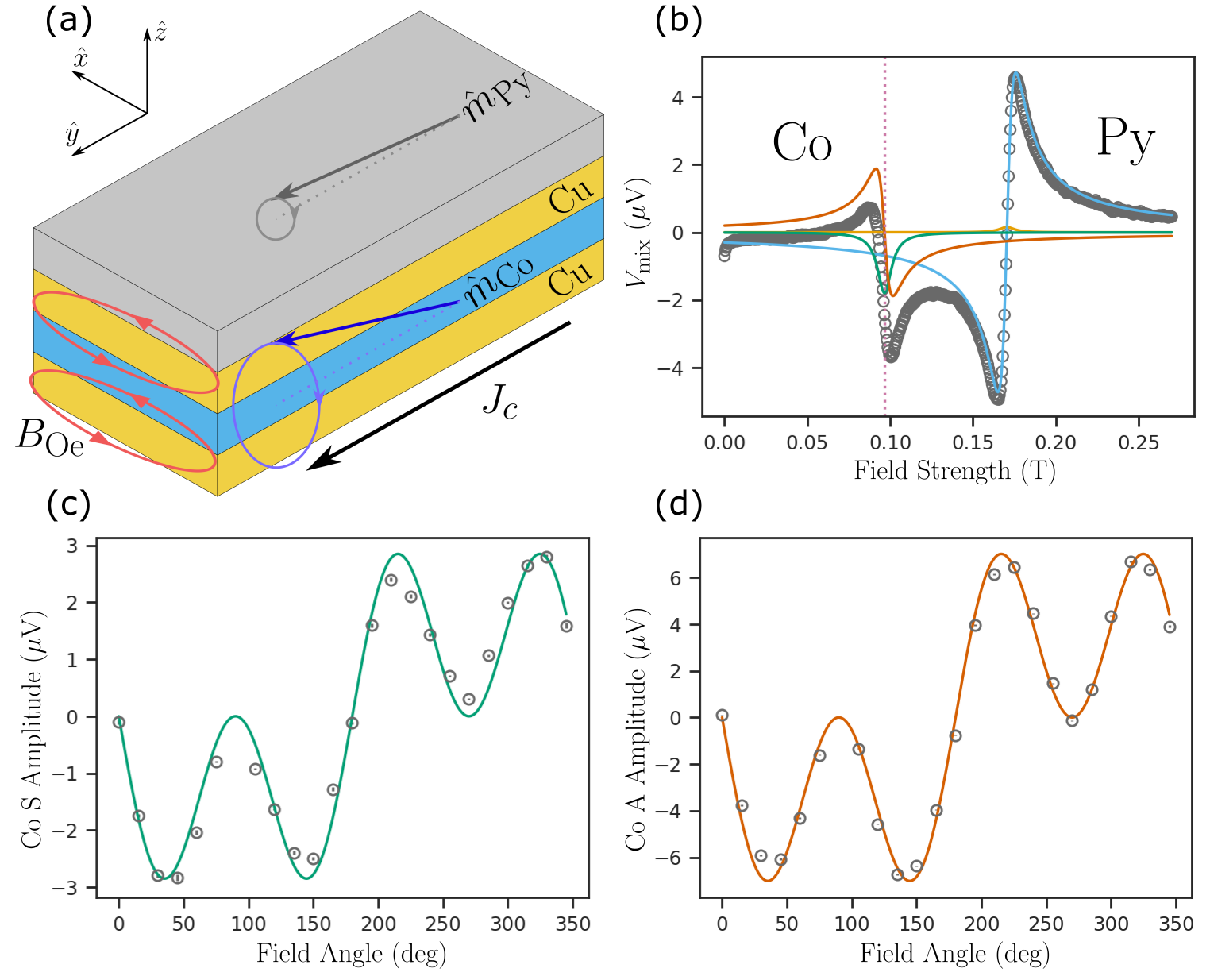}
\caption{(a) Schematic of our device near the Co resonance. The Oersted field from the symmetric spacer layers around the Co will cancel inside of the Co layer. Even at the Co resonance condition the Py precession is still significant. (b) example resonance signal of the $t_\text{Py}=5.7$ nm sample showing the Co (left, lower resonant field) and Py (right, higher resonant field) resonances and their symmetric and antisymmetric components for an applied microwave frequency $f = 8$ GHz. The dashed vertical line represents the Co resonance field. The antisymmetric part of the Py resonance is still substantial there. (c,d) The angular variation in the Co symmetric (c) and antisymmetric (d) resonance components, showing the $\sin(2\phi)\cos(\phi)$ angular dependence expected from SHE-like torques and in-plane Oersted torques, respectively.
}
\label{fig:1}
\end{center}
\end{figure*}

During the measurement, we apply an RF current together with an in-plane external magnetic field at an angle $\phi$ relative to the axis of current flow, and sweep the external field strength to tune through the magnetic resonance (Figure \ref{fig:1}(b)).  
Each sample is studied using several different RF frequencies and field angles $\phi$, and all measurements are performed at room temperature.
We detect a DC signal that arises from mixing between the applied RF current and the resistance variations due to the  angle-dependent magnetoresistance in both magnetic layers, originating from both anisotropic magnetoresistance within the layers and spin Hall magnetoresistance \cite{Nakayama2013}.  
(We work at sufficiently large applied magnetic fields that the magnetizations of the Co and Py layers are saturated parallel in equilibrium, so contributions from giant magnetoresistance are second order in the precession angles, and negligible for the measurement.)
The circuitry of the measurement is described elsewhere \cite{Liu2011, Pai2012, He2018}.  
The two magnetic layers produce two resonances, with the lower resonant field being primarily from the Co and the higher resonant field primarily from the Py.
We fit these resonances to two sets of symmetric and antisymmetric lorentzians to capture the signal voltages from each of the magnetic layers. 

The dependence on the angle $\phi$ for the symmetric and antisymmetric resonance amplitudes of the Co detector layer are shown in Figure \ref{fig:1}(c) and (d):  both follow a $\sin(2\phi)\cos(\phi)$  dependence. 
This is the form expected from  ``conventional" current-induced torques -- from the in-plane component of the Oersted field and from spin-orbit torques arising from a spin current with polarization in-plane and perpendicular to the current.  
It is also the form expected in our geometry for torques arising from interfacial spin rotation as discussed in refs.\ \cite{Amin2018, Baek2018}, and based on the analysis below the same angular dependence will be produced by the torques from dynamic spin pumping. 
We observe no significant contribution from unconventional current-induced torques that would lead to an altered angular dependence \cite{MacNeill2016}.  
We note, however, that since both magnetizations have the same equilibrium orientation for our geometry, our measurements are not sensitive to the reorientable torques arising from the spin anomalous Hall effect or  planar Hall effect \cite{Taniguchi2015, Safranski2018} which create spin currents whose polarization is parallel to the magnetization. 

\section{Analysis: Effect of Dynamic Spin Pumping}\label{sec:dyam_SP}

We can account for the resonant dynamics of the magnetic layers, including the effects of spin pumping between the layers,  using a modified Landau-Lifshitz-Gilbert-Slonczewski (LLGS) equation \cite{Heinrich2003, Pogoryelov2020}
\begin{equation}\label{eq:llgs_coupled}
	\dot{\bm}_i = \alpha_i \bm_i \times \dot{\bm}_i + \btau_{\text{neq},i} + \btau_{\text{eq},i} - \alpha_i' \bm_j\times \dot{\bm}_j
\end{equation}
where $i$ and $j$ identify the magnetic layers, $\btau_{\text{eq},i}$ are the equilibrium torques acting on layer $i$ coming from the applied field and anisotropy, and $\btau_{\text{neq},i}$ are the non-equilibrium torques acting on layer $i$ coming from spin-orbit torques and current-generated Oersted fields. $\alpha_i$ is the effective Gilbert damping parameter for layer $i$, including both the intrinsic damping and the time-averaged effect of the dynamically-pumped spins emerging from layer $i$, while the term containing $\alpha_i'$ describes the effect of the dynamically-pumped spins from layer $j$ impinging on layer $i$ to exert a time-dependent torque. 
Dipole coupling of the form $\bm_i\cdot \bm_j$ is generally also relevant in systems with more than one ferromagnetic layer, but we designed our Cu spacers to be sufficiently thick to minimize this effect, and found by measuring the shift of the minor loop in hysteresis measurements that it is indeed negligible in our system (see Supplementary Information).

Our treatment of the torques due to dynamic spin pumping follow references \cite{Tserkovnyak2005, Heinrich2003, Pogoryelov2020}. Compared to the more-familiar time-averaged DC component of the spin-pumped spin current that generates DC voltage signals in inverse spin Hall experiments \cite{Saitoh2006, Mosendz2010, Azevedo2011} and modifies the magnetic damping \cite{Tserkovnyak2002, Mizukami2002}, the time-varying spin current associated with dynamic spin pumping is much larger.
In systems with just a single ferromagnetic layer the effects from the dynamic portion of the pumped spins are generally time-averaged to zero, but with a second ferromagnetic layer present the pumped spins will induce dynamic coupling between the magnetic layers \cite{Heinrich2003, Richardson2019, Medwal2019, Pogoryelov2020}.
Because the spins pumped by layer $j$ are oscillating and are generally out-of-phase with the precession in layer $i$, they exert a nontrivial torque on layer $i$ which depends on the phase difference between the oscillations of the two layers, introducing field and frequency dependence to the interaction.  Based on this field and frequency dependence, we will show that it is possible to separate the effects of the spin-pumping-induced interaction between the layers from the direct effects of the non-reorientable spin-orbit torque from the source layer.

To model the resonance lineshapes, we consider the case that the resonance field is much larger than the coercive fields of both magnetic layers so that the equilibrium magnetizations of both layers are saturated in the same direction.
We then solve for in-plane ($x$) and out-of-plane ($z$) deflections of each layer, and only keep terms to first order in these deflections from equilibrium.
We can find the deflections of each layer by solving Eq.~(\ref{eq:llgs_coupled}) following the same procedure as \cite{Karimeddiny2020}, noting that we can think of the dynamic spin pumping as effectively modifying the torques on each layer.
After decoupling the in-plane and out-of-plane deflections of each layer, the deflections of the detector layer are
\begin{align}
	&\begin{aligned} 
	m_{dx} = L_d &\left(-\omega_{d2}\left[\tau_{dz} - i\omega \alpha_d' m_{sx}\right]\right. \\
	&\ \left.+ i\omega \left[\tau_{dx} + i\omega \alpha_d' m_{sz}\right]\right)
	\end{aligned}\label{eq:mdx}\\
	 &\begin{aligned}
	m_{dz} = L_d &\left(\omega_{d1}\left[\tau_{dx} + i\omega \alpha_d' m_{sz}\right]\right.\\
	&\ \left.+ i\omega \left[\tau_{dz} + i\omega \alpha_d' m_{sx}\right]\right)
	\end{aligned} \label{eq:mdz}
\end{align}
where the $d$ subscript denotes parameters associated with the detection layer, $L_d = \left[\left(\omega^2-\omega_{d0}^2\right) + i\omega\omega_d^+\alpha_d\right]^{-1}$, $\tau_{dx}$ and $\tau_{dz}$ the in-plane and out-of-plane non-equilibrium torques, $\omega_{d1} = \gamma B$, $\omega_{d2}=\gamma(B + \mu_0 M_{\text{eff},d})$, $\omega_{d0}^2 = \omega_{d1}\omega_{d2}$, and $\mu_0 M_{\text{eff},d} =1.82\pm 0.09$ T for the Co layer based on the precession frequency.
The expressions for the source-layer deflections are the same as the detector layer's but with $s$ and $d$ interchanged and with $\mu_0 M_{\text{eff},s} = 0.93\pm 0.05$ T for the Py.

We can then completely decouple this system by ignoring any terms which are above first order in the $\alpha$'s and $\alpha'$'s.
The result for the in-plane deflection of the detector layer, which generates the signal we are primarily concerned with, is
\begin{widetext}
\begin{equation}\label{eq:mdx_uncoupled}
	\begin{aligned}
	m_{dx} = L_d &\left(-\omega_{d2} \tau_{dz} + \overline{L}_s\alpha_d'\left[(\omega_{s0}^2-\omega^2)(\omega_{s1}+\omega_{d2})\omega^2\tau_{sx}
	- \omega^2\omega_s^+\alpha_s(\omega_{s2}\omega_{d2}+\omega^2)\tau_{sz}\right]\right.\\
	&\ \ \left. +i\omega\left(\tau_{dx} + \overline{L}_s\alpha_d'\left[(\omega_{s0}^2-\omega^2)(\omega_{s2}\omega_{d2}+\omega^2)\tau_{sz}
	+ \omega^2\omega_s^+\alpha_s(\omega_{s1}+\omega_{d2})\tau_{sx}\right]\right)\right)
	\end{aligned}
\end{equation}
\end{widetext}
with $\overline{L}_s = \left[(\omega^2-\omega_{s0}^2)^2 + \omega^2\omega_s^{+2}\alpha_s^2\right]^{-1}$ and $\omega_{i}^+ = \omega_{i1}+\omega_{i2}$.
The real part of the expression within the large parentheses corresponds to the antisymmetric component of the detector-layer resonance and the imaginary part to the symmetric component.  The terms proportional to $\overline{L}_s \alpha_d'$ represent modifications of the usual expression for the ST-FMR resonance caused by spin pumping from the source layer.  Through these terms, torques acting on the source layer ($\tau_{sx}$, $\tau_{sz}$) produce a modification of the torques acting on the detector layer, with the strength of the effect depending on the amplitude of the source-layer resonance through $\overline{L}_s$.
The terms in Eq.~(\ref{eq:mdx_uncoupled}) inside the square brackets containing $(\omega_{s0}^2-\omega^2)$ correspond to the antisymmetric component of the source-layer resonance while the other terms correspond to the symmetric component.

\begin{figure*}
\begin{center}
\includegraphics[width=\linewidth]{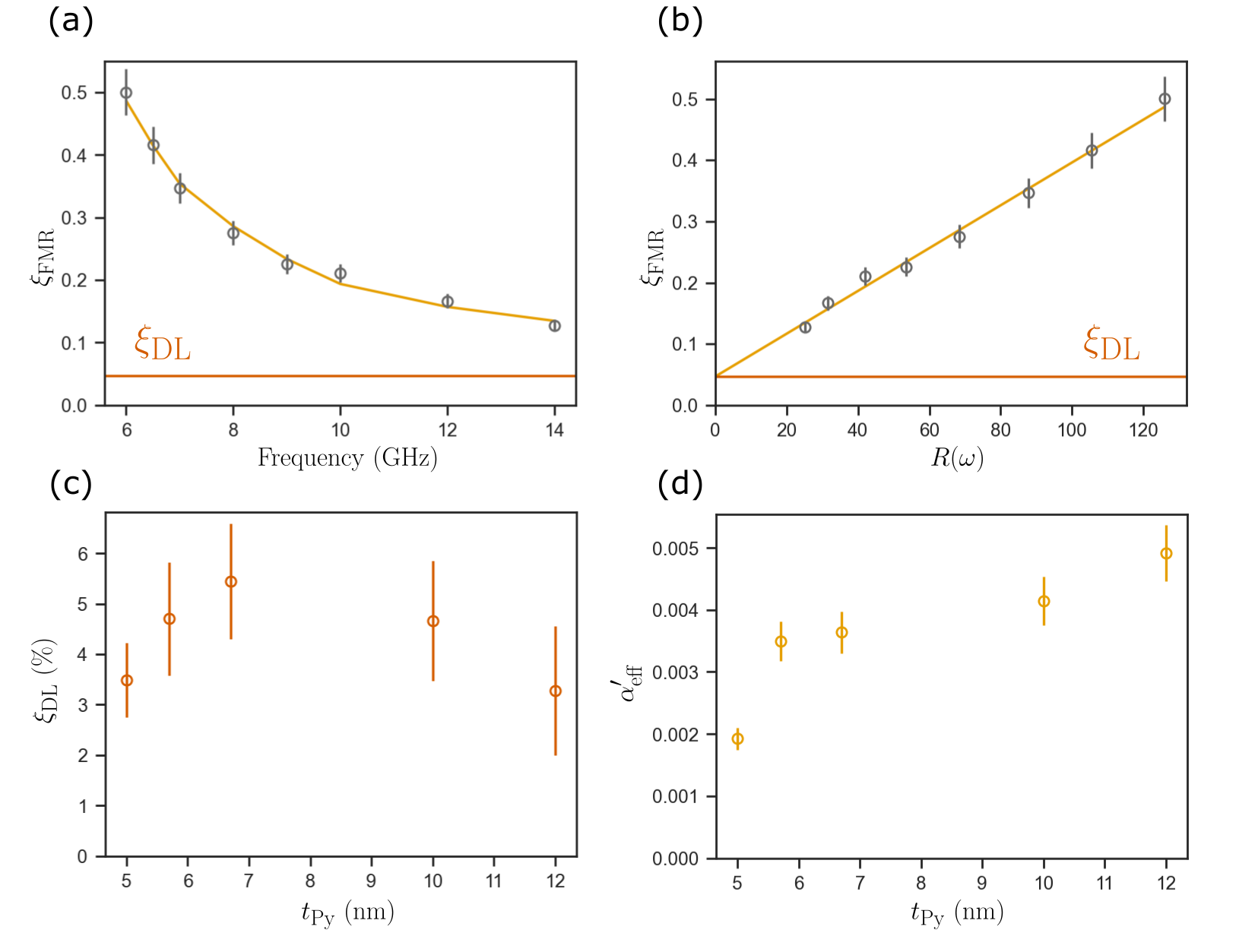}
\caption{(a,b) Measured values of $\xiFMR$ for a sample with 5.7 nm Py with a fit to Eq.~(\ref{eq:fiteq}). In (a) we plot the data versus frequency and in (b) we plot the same data versus calculated values of $R(\omega)$ to show that the data vary linearly with $R(\omega)$. The horizontal line is the extracted value of $\xi_\text{DL}$, which is the asymptote of (a)/the intercept of (b). (c) The effect of Py thickness on $\xi_\text{DL}$. (d) The dependence of the dynamic spin pumping parameter on the source Py layer thickness. 
}
\label{fig:2}
\end{center}
\end{figure*}

This expression can be simplified significantly for application to our measurements.   
First, the symmetric part of our source (Py) resonance is essentially zero near the detector (Co) resonance field as seen in Figure \ref{fig:1}(b), so we can ignore all terms in  Eq.~(\ref{eq:mdx_uncoupled}) which correspond to the symmetric part of the source resonance.
Second, the dampinglike torque acting on the source layer, $\tau_{sx}$, is quite small compared to the Oersted torque on the detector layer, $\tau_{dz}$, as illustrated in Figure \ref{fig:1}(b) by the fact that the symmetric part of the Py resonance is tiny compared to the antisymmetric part of the Co resonance. 
This allows us to ignore all terms in Eq.~(\ref{eq:mdx_uncoupled}) which include $\tau_{sx}$ as well.
After these simplifications, the in-plane oscillation of the detector layer takes the form
\begin{equation}\label{eq:mdx_uncoupled_simple}
	\begin{aligned}
	m_{dx} &= L_d \left(-\omega_{d2} \tau_{dz} \right.\\
	&\left.+i\omega \left(\tau_{dx}+ \overline{L}_s\alpha_d'(\omega_{s0}^2-\omega^2)(\omega_{s2}\omega_{d2}+\omega^2)\tau_{sz}\right)\right).
	\end{aligned}
\end{equation}

The amplitude of the DC mixing signal we detect has the form
\begin{equation}\label{eq:mixingvoltage}
	V_\text{mix} = \frac{I}{2}\left(R_d^\text{AMR}\Re{m_{dx}}
	+ R_s^\text{AMR}\Re{m_{sx}}\right)\sin(2\phi),
\end{equation}
where $I$ is the total current flowing in the multilayer and $R_i^\text{AMR}\sin(2\phi)$ is the $\phi$-derivative of the resistance change of the whole heterostructure due to the angle-dependent magnetoresistance of layer $i$.
In general, it is also necessary to include an additional voltage signal due the DC component of the pumped spin current and the inverse spin Hall effect \cite{Saitoh2006, Mosendz2010, Azevedo2011}, but we calculate that this contribution is negligible for our samples (see Appendix A).

Even at the position of the detector-layer resonance it is not correct to ignore the contribution proportional to the source-layer magnetoresistance (the second term in Eq.~(\ref{eq:mixingvoltage})), because precession of the detector layer will produce dynamic spin pumping that drives precession of the source layer.
Since this precession of the source layer is driven by spin pumping from the detector layer, near the detector layer's resonance condition the source-layer precession has the same resonant field and linewidth as the detector-layer resonance.
This means that when we measure the lineshape which we attribute to the detector layer, some of this signal is actually due to the source-layer precession and the mixing voltage which results.

We can take into account the source-layer oscillations by writing their in-plane component in the form
\begin{equation}\label{eq:msx_near_detection}
    \begin{aligned}
    m_{sx} = &L_s(-\omega_{s2}\tau_{sz} + i\omega \tau_{sx})\\
    &\begin{aligned}
    + L_d\overline{L}_{s}(\omega_{s0}^2-\omega^2)&\left[\omega^2(\omega_{d1}+\omega_{s2})\tau_{dx}\alpha_s'\right.\\
    &\ \left.+ i\omega(\omega_{d2}\omega_{s2}+\omega^2)\tau_{dz}\alpha_s'\right]
    \end{aligned}
    \end{aligned}
\end{equation}
where the first term is the source-layer resonance in the absence of any dynamic spin pumping coupling and the $L_d$ term the one producing a resonance in the source layer of the same shape as the detector layer.
Near the detector-layer resonance condition, the first term is essentially constant and hence will not play a significant role in our analysis. 

Combining Eqs.~(\ref{eq:mdx_uncoupled_simple}), (\ref{eq:mixingvoltage}) and (\ref{eq:msx_near_detection}), the voltage amplitudes we measure upon fitting the detector-layer resonance as a sum of symmetric and antisymmetric Lorentzians is
\begin{widetext}
\begin{align}
    V_{dS} &= \frac{I R_d^\text{AMR}\sin(2\phi)}{2}\frac{1}{\alpha_d\omega_d^+}\left[\tau_{dx}+\overline{L}_{sA}(\omega_{s2}\omega_{d2}+\omega^2)\tau_{sz}\alpha_d'\right] + \frac{I R_s^\text{AMR}\sin(2\phi)}{2}\frac{1}{\alpha_d\omega_d^+}\overline{L}_{sA}(\omega_{s2}\omega_{d2}+\omega^2)\tau_{dz}\alpha_s'\label{eq:VdS_full}\\
    V_{dA} &= \frac{I R_d^\text{AMR}\sin(2\phi)}{2} \frac{1}{\alpha_d\omega_d^+}\frac{\omega_{d2}}{\omega}\tau_{dz}\label{eq:VdA_full}
\end{align}
\end{widetext}
where $\overline{L}_{sA} = \overline{L}_s (\omega_{s0}^2-\omega^2)$.
The first part of each mixing voltage comes from the detector layer and the second part is due to the source layer.
There is also a small contribution of the source-layer resonance to $V_{dA}$ that is not included in Eq.\ (\ref{eq:VdA_full}) because it contributes no more than $1\, \%$ of signal from the detector layer for our system, so we can ignore it in our analysis.

We are left with two significant modifications to the ST-FMR signals caused by dynamic spin pumping, both of which modify the measured symmetric lineshape amplitude of the detector layer. 
The first is due to spins pumped from the source layer to the detector layer, thereby altering the detector layer resonance, an effect which is proportional to the Oersted torque on the source layer, $\tau_{sz}$, with an amplitude proportional to the antisymmetric component of the source-layer resonance.
This contribution adds directly to the dampinglike torque on the detector layer, $\tau_{dx}$.
The second is due to spins pumped from the detector layer to the source layer, causing the source layer to oscillate with the same general lineshape as the detector-layer resonance. The size of this term proportional to the Oersted torque on the detector layer $\tau_{sz}$, and it also has an amplitude proportional to the antisymmetric component of the source-layer resonance.
From Eqs.\ (\ref{eq:VdS_full}) and (\ref{eq:VdA_full}) it is clear that the spin-pumping terms are not expected to alter the usual angular dependence of the ST-FMR signals ($\propto \sin(2\phi)\cos(\phi)$) as long as the current-induced torques are proportional to $\cos(\phi)$, the dependence expected for both conventional-symmetry spin-orbit torques and Oersted torques \cite{MacNeill2016}.

The efficiency of the antidamping spin-orbit torque generated by the Py source layer acting on the Co detector layer can be characterized by the antidamping torque efficiency, defined as
\begin{equation}
    \xi_\text{DL} = \tau_{dx} \frac{e M_\text{sat,d}t_d}{\mu_B J_{c,s}},
\end{equation}
where $e$ is the electron charge, $M_{\text{sat},d} = 1.19\pm 0.08 \times 10^6$ A/m for the Co layer, $t_d = 8$ nm for the Co layer, and $J_{c,s}$ the charge current density within the source magnetic layer.
This efficiency, and also the influence of the dynamic spin pumping, can be determined by analyzing how the ratio of the symmetric and antisymmetric resonance amplitudes depends on measurement frequency.  Following ref.~\cite{Pai2015}, we first define an intermediate parameter 
\begin{equation}\label{eq:xiFMR}
    \xi_{\text{FMR}}=\frac{e\mu_0 M_{\text{sat},d}t_st_d}{\hbar}\sqrt{\frac{\omega_{d2}}{\omega_{d1}}}\frac{V_{dS}}{V_{dA}}
\end{equation}
where $t_s=t_\text{Py}$.  
The values of $\xi_{\text{FMR}}$ extracted for a sample with $t_s= 5.7$ nm as a function of different applied RF frequencies are shown by the circles in Figure~\ref{fig:2}(a).  
We find that $\xi_{\text{FMR}}$ increases by more than a factor of 3 as the frequency is decreased from 14 GHz to 6 GHz, a much stronger dependence than is found in samples with non-magnetic source layers \cite{Liu2011}. 
We will show that this can be explained by dynamic spin pumping, in that for lower RF frequencies the resonance fields of the two magnetic layers move closer together, increasing the source layer resonance amplitude, and hence the value of $\overline{L}_{sA}$, at the detector-layer resonance condition which enhances the effects of the dynamic spin pumping.

For magnetic layers as thick as those we employ, $\geq 5$ nm, and since the detector layer's interfaces are symmetric and are with the light metal Cu, Oersted torques should dominate over any weak interfacial field-like torques for both the source and detector layers \cite{Miron2011, Pai2015, Jiang2020}.
We therefore assume that $\tau_{dz}=\gamma\mu_0J_{c,s}t_s/2$ (only the current density within the source layer contributes to the Oersted field acting on the detector layer because the contributions from the Cu spacer layers cancel and the Ta layers have high resistivity and negligible current densities) and $\tau_{sz} = \gamma \mu_0 (J_{c,d}t_d+J_{c,\text{Cu}} t_\text{Cu})/2$ (the current creating the Oersted field in the source layer comes from the detector layer and the Cu layers, which comprise the rest of the conductive layers in the device). 
The ratio of these torques can be written $\tau_{sz}/\tau_{dz}=-(1-x_s)/x_s$ where $x_s$, the fraction of the total RF current flowing within the source layer, is in the range of 0.06 to 0.12 depending on the Py thickness and is  calculated from the layer resistivities using a parallel-conduction model (see Supplementary Information).  
It then follows that
\begin{equation}\label{eq:fiteq}
    \xi_{\text{FMR}}= \xi_\text{DL} + R(\omega)\alpha'_\text{eff}
\end{equation}
with
\begin{align}
    R(\omega) &= - \frac{e\mu_0 M_{\text{sat},d}t_st_d}{\hbar}\overline{L}_{sA}(\omega_{s2}\omega_{d2}+\omega^2)\frac{1-x_s}{x_s} \label{eq:fit_freq}\\
    \alpha'_\text{eff} &= \alpha_d' - \frac{x_s}{1-x_s}\frac{R_s^\text{AMR}}{R_d^\text{AMR}} \alpha_s'. \label{eq:fit_damping}
\end{align}
All of the parameters that are frequency dependent are contained within $R(\omega)$. Furthermore, all of the parameters that enter $R(\omega)$ are independently measurable, so we can use Eq.~(\ref{eq:fiteq}) to fit to the data in Figure~\ref{fig:2}(a), using just two adjustable fit parameters, $\xi_\text{DL}$ and $\alpha'_\text{eff}$.  The fit is shown both in Figure~\ref{fig:2}(a) and in Figure~\ref{fig:2}(b), which display the same data points plotted as a function of measurement frequency in (a) and as a function of $R(\omega)$ in (b).  For the $t_\text{Py} = 5.7$ nm sample, we find $\xi_\text{DL} = 0.05\pm 0.01$ and $\alpha'_\text{eff} = 4.0\pm 0.4 \times 10^{-3}$.  The dominant source of uncertainty comes from the determination of $M_{\text{eff},d}$ which we measure by fitting the frequency dependence of the detector-layer resonant field to the Kittel equation.  
We note in samples without dynamic spin pumping, and in which like our samples the field-like spin-orbit torque is negligible relative to the Oersted torque, $\xi_\text{DL} = \xi_{\text{FMR}}$.  Therefore, Figure~\ref{fig:2}(a) illustrates that the neglect of spin-pumping in our analysis would lead to a large overestimate of $\xi_\text{DL}$.

We have performed the same analysis on samples in which the Py layers vary in thickness from 5 nm to 12 nm.  
The results for  $\xi_\text{DL}$ and $\alpha'_\text{eff}$ are shown in Figure~2(c) and (d).
The overall dampinglike torque efficiency $\xi_{\text{DL}}$ of the magnetization-independent spin-orbit torque from Py through the 8 nm Cu spacer to the Py layer is $0.04 \pm 0.01$, and it does not have a significant dependence on the Py thickness within our experimental uncertainty.  
The value we determine for $\alpha'_\text{eff}$ does depend on the Py thickness, changing by more than a factor of 2 between Py thicknesses of 5 and 12 nm.  
The parameter $\alpha'_d$ should not depend on the Py thickness, so the changes in $\alpha'_\text{eff}(t_\text{Py})$ indicate that the second term in Eq.~(\ref{eq:fit_damping}) is important as well as the first -- that the signal at the detector resonance is affected by spin pumping from the detector layer to the source layer that excites precession in the source (Py) layer.  
We are not able to make a separate determination of the parameters $\alpha'_d$ and $\alpha'_s$ in Eq.~(\ref{eq:fit_damping}), however, because we are not able to make an accurate calibration of the ratio $R_s^\text{AMR}/R_d^\text{AMR}$.  
We can calibrate the anisotropic magnetoresistance of individual Co and Py layers, but we find that when they are combined within a trilayer there is also a significant additional contribution from spin Hall magnetoresistance that cannot be calibrated without separate control over the magnetization angles in the two layers.

\section{Discussion}

We can compare our results to previous measurements.
Efforts to determine the non-reorientable spin Hall effect in ferromagnetic layers via spin Seebeck and spin pumping experiments are complicated by difficulties in determining the interface spin mixing conductance \cite{Zhu2019c} and the spin diffusion length.
Furthermore, many of these experiments do not provide sufficient information to distinguish between the reorientable and non-reorientable components of the spin current generated by Py.
The two measurements with direct quantitative claims we have found in this class of experiments give a value for the non-reorientable spin Hall ratio within Py of $\theta_\text{SH}(\text{Py})$ = 0.02 by spin pumping in YIG/Cu/Py \cite{Wang2014, Du2014}.
The quantity we measure, the spin-orbit torque efficiency, $\xi_\text{DL}$, is related to the spin Hall ratio as $\xi_\text{DL} = \theta_\text{SH}(\text{Py})T_\text{int}$ where $T_\text{int}$ is an interfacial spin transfer coefficient less than or equal to 1. 
Therefore our measurements establish a lower bound for the non-reorientable spin Hall effect for the Py in our samples, $\theta_\text{SH}(\text{Py}) \ge 0.04 \pm 0.01$. 
We suspect the difference is that the spin mixing conductance may be overestimated \cite{Zhu2019c} in refs.\ \cite{Wang2014, Du2014}, leading to an underestimate of $\theta_\text{SH}(\text{Py})$.  
Miao et al.\ \cite{Miao2013} and Wu et al.\ \cite{Wu2015} studied YIG/Py samples and quoted values of $\theta_\text{SH}(\text{Py})$ relative to values for similar samples made with Pt instead of Py: $\theta_\text{SH}(\text{Py})/\theta_\text{SH}(\text{Pt}) =$  0.38 \cite{Miao2013} and 0.98 \cite{Wu2015}. 
These estimates rely on assumptions that the interfacial spin transparency and spin diffusion lengths are similar for YIG/Py and YIG/Pt.

Das et al.\ \cite{Das2017, Das2018} detected both reorientable and non-reorientable components of spin-orbit torque generated by Py using nonlocal measurement scheme where a spin current generated in Py was converted to magnons in YIG which then were converted back into a spin current in Pt or Py and transduced into a voltage through the inverse spin Hall effect.
Their value of $\theta_\text{SH}(\text{Py})$ is only given in relation to that of Pt, they quote $[G_\text{Py} \theta_\text{SH}(\text{Py})]/[G_\text{Pt} \theta_\text{SH}(\text{Pt})] = 0.09$ where $G_i$ is a parameter describing the spin current to magnon conversion in each of the materials.
Yang et al.\ \cite{Yang2020} attempted to measure the non-reorientable component of the spin curernt alone from Py using lineshape analysis of ST-FMR signals for Py/YIG and CoFeB/YIG samples (both with no spacer layer), but without accounting for dynamics spin pumping. 
They claimed values of the non-reorientable spin-orbit torque efficiencies $\xi_\text{DL}= 0.009$ for Py acting on YIG and $-0.0014$ for CoFeB acting on YIG; values much lower in magnitude than our result. 
We suspect that these low values may be due primarily to poor interfacial spin transmission from the Py into the YIG.  
Low spin transparency has also been a general feature found in spin-orbit torque measurements from heavy metals acting on iron garnets (see the comparisons by Gupta et al.\ \cite{Gupta2020}).

Based on measurements of conventional FMR in magnet/spacer/magnet samples, the parameter $\alpha'_\text{eff}$ is expected to have a scale comparable to the intrinsic Gilbert damping in magnetic layers, $\alpha$ \cite{Heinrich2003}.  This is true for our samples. The values of $\alpha'_\text{eff}$ plotted in Fig.~\ref{fig:2}(d) are about half the damping parameters in our Co and Py layers determined by the ST-FMR fits.

\section{Conclusion}\label{sec:conclusion}

We have measured and analyzed spin-torque ferromagnetic resonance (ST-FMR) signals for samples containing the layer structure Py source layer/Cu spacer/Co detector layer. 
We find that the resonance amplitudes and lineshapes are determined by a combination of direct current-induced torques and dynamic spin pumping between the magnetic layers.  In fact, the strength of the torque created by dynamic spin pumping can be substantially larger than the direct current-induced spin-orbit torque, so the contribution from dynamics spin pumping should be considered whenever analyzing resonant measurements in samples with more than one ferromagnetic layer \cite{Yang2020,Keller2019}. The frequency dependence of the ST-FMR signals allows these two effects to be separated, and therefore our experiment provides independent measurements of the non-reorientable spin-orbit torque generated by the Py layer (the portion of the spin-orbit torque that does not depend on the orientation of the Py magnetization) and the strength of dynamical spin pumping.  We find an efficiency for the non-reorientable spin-orbit torque generated by Py (and acting through a 8 nm Cu spacer) of $0.04 \pm 0.01$. The strength of the dynamic spin pumping, as characterized by the parameter $\alpha'_\text{eff}$, appears consistent with previous measurements using conventional ferromagnetic resonance \cite{Heinrich2003}. 

\section{Acknowledgements}

We thank Bob Buhrman for discussions.  This research was supported in part by Task 2776.047 of ASCENT, one of six centers in JUMP, a Semiconductor Research Corporation program sponsored by DARPA, and by the NSF (DMR-1708499). 
The devices were fabricated using the shared facilities of the Cornell NanoScale Facility, a member of the National Nanotechnology  Coordinated Infrastructure (NNCI), and the Cornell Center for Materials Research, both of which are supported by the NSF (Grants No. NNCI-1542081 and No. DMR-1719875).

\appendix

\section{DC Spin Pumping Contribution}\label{apx:DC_SP}

\begin{figure}
\includegraphics[width=3.2in]{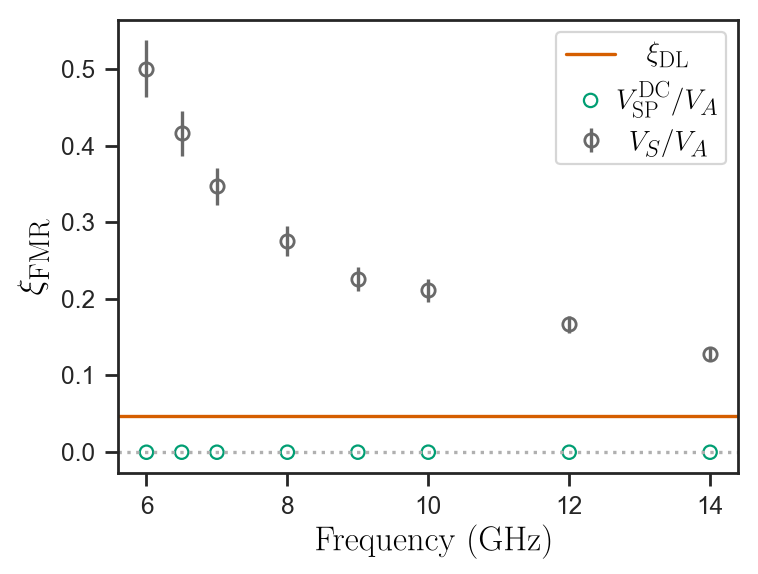}
\caption{Comparison of the DC spin pumping contribution to $\xiFMR$ to the total signal for 5.7 nm Py. The contribution is so small as to be negligible.}
\label{fig:A1}
\centering
\end{figure}

The DC component of the pumped spin current from the Co is converted into a DC charge current through the inverse spin Hall effect in the Py layer, as described in ref. \cite{Karimeddiny2020}.
The generated DC spin pumping voltage follows the form
\begin{widetext}
\begin{equation}
    V_{\text{SP}}^\text{DC} = -\frac{2e}{\hbar}\theta_\text{SH}R_\text{tot} W \sin\phi \frac{\hbar}{4\pi}g_\text{eff}^{\uparrow\downarrow}\lambda_\text{sd,s}\left[\frac{\omega_{d1}\tau_{dx}^2 +\omega_{d2}\tau_{dz}^2}{(\alpha_d\omega_d^+)^2} L_{dS}(B)\right]\tanh\left(\frac{t_s}{2\lambda_{sd,s}}\right)
\end{equation}
\end{widetext}
where $R_{\text{tot}}$ is the total resistance of the heterostructure, $W$ is the width of the device, $\lambda_{\text{sd},s}$ is the spin diffusion length of the source, and $L_{dS}(B)$ is the symmetric lineshape of the detector layer.
We have either measured or can estimate with reasonable certainty all quantities in this expression.
For the $t_s = 5.7$ nm Py sample, $R_\text{tot} = 12.1\, \Omega$, $W=20\, \mu$m, $\alpha_d = 0.014$, $\theta_\text{SH}\approx 0.05$, we approximate $g_\text{eff}^{\uparrow\downarrow}\approx 8$ nm$^{-2}$ \cite{Zhu2019c} and $\lambda_{\text{sd},s} \approx 10$ nm \cite{Bass2016} as reasonable estimates compared to other systems, we approximate 1.5 mA of RF current is flowing in the Py layer using the input microwave power after accounting for loss in the cabling and current shunting, which leads to $\tau_{dx}\approx 4\times 10^{6}$ s$^{-1}$ and $\tau_{dz}\approx 8\times 10^{6}$ s$^{-1}$, and $\omega_{d1},\ \omega_{d2}$ and $\omega_d^+$, which are evaluated at the Co resonant field and hence depend on frequency, are on the order of 10 GHz, 300 GHz and 300 GHz, respectively.
To compare this more directly to what we have measured before, we utilize the fact that this will only add to the symmetric component of the resonance.
Since we look at the ratio of the symmetric and antisymmetric resonance amplitudes, we compute $V_{\text{SP}}^\text{DC}/V_A$ where $V_A$ is the amplitude of the antisymmetric component of the Lorentzians.
We then compare this to $V_S/V_A$ in Figure~\ref{fig:A1}, where we have included constants to convert the voltage ratios to torque efficiencies.
This contribution ends up being negligibly small, compared to both the total signal due to dynamic spin pumping and the signal due to the dampinglike torque from the source layer, never rising above $5\, \%$ of the measured symmetric signal for the thickest Py layers and at the highest frequencies but is less than $1\, \%$ for most thicknesses and frequencies.
We therefore are justified in ignoring this contribution in our main analysis.

\bibliography{2020_trilayer_refs.bib}

\newpage
\onecolumngrid
\section{Evidence that Bilinear Coupling is Negligible in Our Samples}\label{sec:bilinear}

To test whether we have any significant dipole or other bilinear coupling of the form $\bm_1\cdot \bm_2$ between the two magnetic layers, we conducted vibrating sample magnetometry (VSM) measurements of the magnetization hysteresis loop for the bilayer.
We performed a total of three hysteresis sweeps for each sample.  The first covers a field range larger than the coercive field for both magnetic layers and allows us to read the full hyseresis curve of the bilayer.
The other two are minor loops which only switch the magnetization of one layer while keeping the magnetization of the other layer constant.
The results of these hysteresis sweeps on a device with a 6.7 nm Py layer are shown in Supplementary Fig.~\ref{fig:S1}.

The difference in the center of the minor hysteresis loops should be equal to the bilinear coupling effective field between the two layers, since the magnet with the larger coercive field has a different orientation and hence would induce opposite directions of the bilinear exchange field in the two minor loops.
The difference in the minor loop centers is only 2 Oe, and is similar for all of the other Py thicknesses.  This is sufficiently weak coupling that it can be neglected in our analysis relative to the coupling produced by dynamic spin pumping.

\begin{figure*}[h]
\begin{center}
\includegraphics[width=\linewidth]{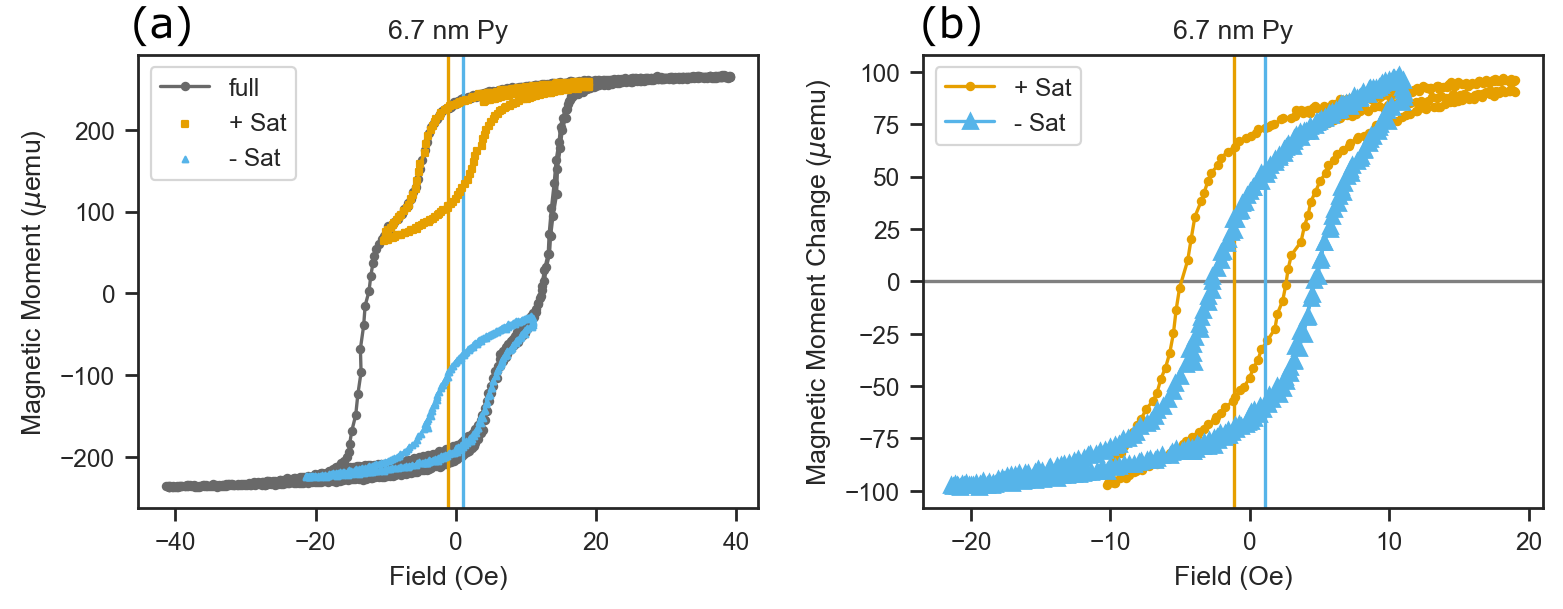}
\caption{(a) The full hysteresis curve with the two minor loops superimposed for the 6.7 nm Py sample. The vertical lines indicate the center of the minor hysteresis loops. (b) The minor hysteresis loops superimposed. The centers of the minor loops are 2 Oe apart, indicating that there is not significant bilinear coupling.
}
\label{fig:S1}
\end{center}
\end{figure*}

\onecolumngrid
\section{Resistivities of the Ferromagnetic Layers}\label{sec:resistivity}

To measure the resistance of the ferromagnetic layers in our devices, we use a separate series of samples with the structure Ta (1) / Cu (3) / FM ($t_\text{FM}$) / Ta (1), which were created so that there would be less shunting from the ferromagnetic layers leading to a better signal to noise ratio.
By investigating the dependence of the heterostructure's resistance on the ferromagnetic layer's thickness, we are able to estimate the resistivity of the ferromagnetic layer.
In a simple parallel conductor model, we would expect the heterostructure resistance to have a linear dependence on the inverse ferromagnetic layer thickness, but we found that since the thickness of our ferromagnetic layers are generally comparable to the mean free path of the ferromagnets, a full Fuchs-Sondheimer model \cite{Warkusz1978} was needed to get reasonable results.
Using electron mean free paths from the literature \cite{Gall2016} for Co and Ni (which we used as an approximation for the mean free path of Py), we were able to find the bulk resistivities of the Co and Py layers using the Fuchs-Sondheimer model to fit the thickness dependence of the device conductivity as shown in Supplementary Figure~\ref{fig:S3}.
We found the bulk resistivity of Co to be $16.2\pm 0.2$ $\mu\Omega$-cm and the bulk resistivity of Py to be $24.9\pm 0.4$ $\mu\Omega$-cm. 
When calculating the shunting factors for the main text, we calculated the resistances of the various layers using the full Fuchs-Sondheimer model for that particular thickness of the ferromagnetic layer.

\begin{figure*}
\begin{center}
\includegraphics[width=\linewidth]{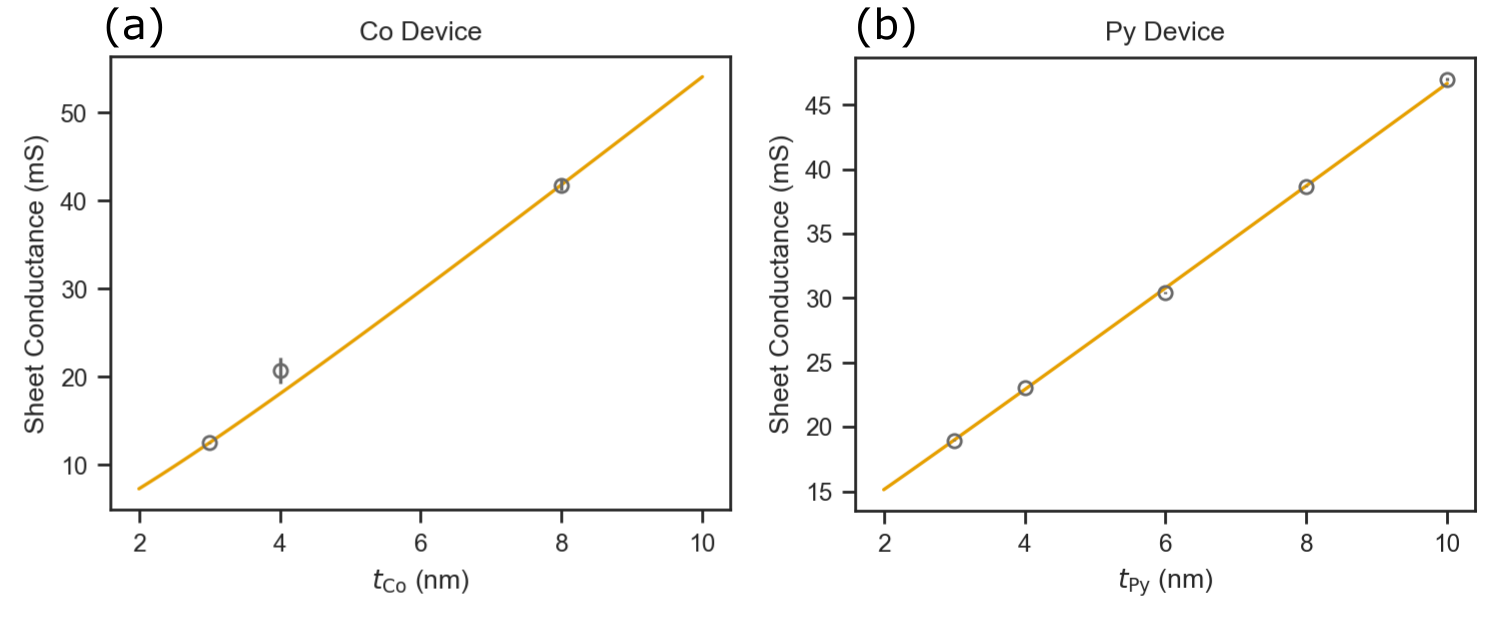}
\caption{Conductance of (a) Co and (b) Py devices with a fit to the Fuchs-Sondheimer model to determine the resistivity of each layer.
}
\label{fig:S3}
\end{center}
\end{figure*}

\section{Demonstration that Both Anisotropic Magnetoresistance and Spin Hall Magnetoresistance Contribute to the Angle-Dependent Magnetoresistance of the Magnetic Trilayers}\label{sec:AMR}

To measure the anisotropic magnetoresistance (AMR) in our devices, we use a separate series of samples with the structure Ta (1) / Cu (3) / FM ($t_\text{FM}$) / Ta (1), which were created so that there would be less shunting from the ferromagnetic layers leading to a better signal-to-noise ratio. 
We measure the resistance of our device with a nanovoltmeter while rotating a magnetic field in the plane.
We measure the resistance change of the whole device, but we would like to measure the resistance change of just the ferromagnetic layer.
To get this quantity, we use a parallel resistor model and assume that the AMR resistance change is small compared to the resistance of the device.
We define $\phi$ as the angle between the current and external magnetic field, $R_\text{tot}(\phi)$ the total device resistance, $R_\text{NM}$ the resistance of the non-magnetic layer, and $R_\text{FM}(\phi) = R_0 + R^{\text{AMR},0} \cos^2 \phi$ the ferromagnet resistance with $R_0$ the base resistance and $R^{\text{AMR},0}$ the AMR resistance.
Expanding, we find that
\begin{align}
    \frac{1}{R_\text{tot}(\phi)} &= \frac{1}{R_\text{NM}} + \frac{1}{R_\text{FM}(\phi)}\\
    &= \frac{1}{R_\text{NM}} + \frac{1}{R_0} \frac{1}{1 + R^{\text{AMR},0}/R_0 \cos^2\phi}\\
    \frac{1}{R_\text{tot}(\phi)} &\approx \frac{1}{R_\text{NM}} + \frac{1}{R_0} - \frac{R^{\text{AMR},0}}{R_0^2}\cos^2\phi
\end{align}
which provides a way for us to extract the desired quantity from fitting the total conductance as a function of applied field angle.
Fits for the devices with 10 nm Py and 8 nm Co are shown in Supplementary Fig.~\ref{fig:S2}.

\begin{figure*}
\begin{center}
\includegraphics[width=\linewidth]{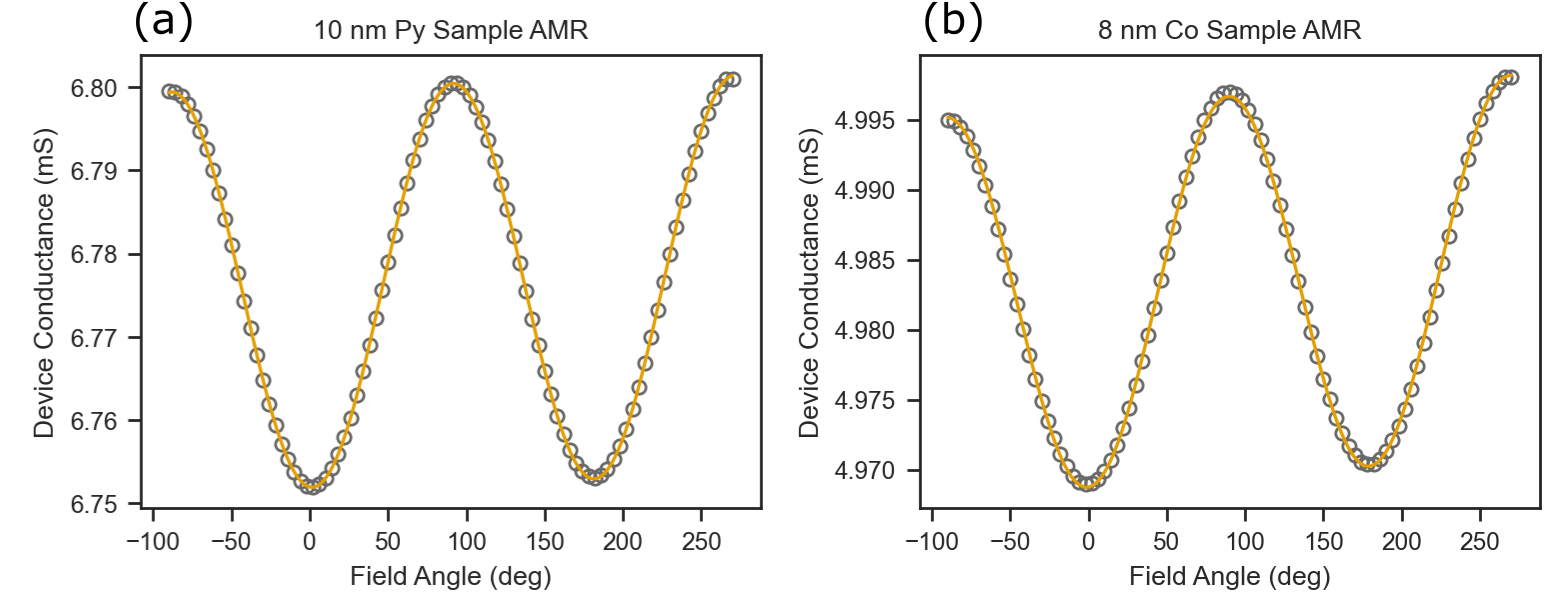}
\caption{Device conductance as a function of applied field angle for (a) the 10 nm Py device and (b) the 8 nm Co device.
}
\label{fig:S2}
\end{center}
\end{figure*}

We have also measured the AMR resistance change of a full trilayer structure, Cu (8) / Co (8) / Cu (8) / Py ($t_\text{Py}$=10) from the main text.
Using the values we have measured above for the 8 nm Co and 10 nm Py single layers, we would expect the conductance of this device to vary by 75 $\mu$S due to the AMR by adding the contributions from each of the ferromagnetic layers.
However, we actually measure a conductance variation of 23 $\mu$S, which is over three times smaller than what we expect from the single-layer measurements.
We attribute this discrepancy to the presence of spin Hall magnetoresistance in the full trilayer samples, which only exists when we have layers with non-negligible spin Hall effect.
In the single layers, the only other conductive layer was the Cu which has a negligible spin Hall effect.
This has precluded a simple measurement of the appropriate $R^\text{AMR}_i$ to use in the expressions in the main text.

\section{Derivation of the Effects of Dynamic Spin Pumping}\label{sec:dynamSP}

To derive the effect which the dynamic spin pumping has on the resonances of our layers, we begin with the modified LLGS equation proposed by Heinrich \cite{Heinrich2003}
\begin{equation}
	\dot{\bm}_i = \alpha_i^0 \bm_i \times \dot{\bm}_i + \btau_{\text{neq},i} + \btau_{\text{eq},i} + \alpha_i'(\bm_i\times \dot{\bm}_i - \bm_j\times \dot{\bm}_j)
\end{equation}
where $\btau_{\text{neq},i}$ are the nonequilibrium torques which oscillate at the RF frequency such as the Oersted torque or the spin-orbit torque, $\btau_{\text{eq},i}$ are the equilibrium torques which do not oscillate such as those from the static external field or from anisotropy, $\alpha_i^0$ is the intrinsic damping parameter and $\alpha_i'$ is a parameter describing the effect of spin pumping that both layers have on layer $i$.
We can combine the effect of damping with the spin pumping out of layer $i$ to arrive at a modified damping parameter $\alpha_i = \alpha_i^0 + \alpha_i'$
\begin{equation}
\dot{\bm}_i = \alpha_i \bm_i \times \dot{\bm}_i + \btau_{\text{neq},i} + \btau_{\text{eq},i} - \alpha_i' \bm_j\times \dot{\bm}_j.
\end{equation}
First we assume that both layers are saturated in the same direction and have the same equilibrium magnetization direction. 
We use a coordinate system where $\hat{y}$ is along the equilibrium direction of the magnetization and $\hat{z}$ is normal to the film. 
The magnetization is essentially constant along the $\hat{y}$ axis, so our problem reduces to a 2D problem in the $x-z$ plane. 
The torques we need to consider are
\begin{equation}
    \btau_{\text{eq},i} = \left(
    \begin{matrix}
    m_{iz} \omega_{i2}\\
    -m_{ix} \omega_{i1}
    \end{matrix}\right),\ \ \ \ \btau_{\text{neq},i} = \left(
    \begin{matrix}
    \tau_{ix}\\
    \tau_{iz}
    \end{matrix}\right),\ \ \ \ \bm_i\times\dot{\bm}_i = \left(
    \begin{matrix}
    \dot{m}_{iz}\\
    -\dot{m}_{ix}
    \end{matrix}\right)
\end{equation}
with $\omega_{i1} = \gamma B$, $\omega_{i2} = \gamma\left(B + \mu_0 M_{\text{eff},i}\right)$, $\gamma$ the gyromagnetic ratio and $M_{\text{eff},i}$ the effective magnetization of layer $i$.
Putting this all into the modified LLGS, arrive at
\begin{align}
	\left(
	\begin{matrix}
	\dot{m}_{sx}\\
	\dot{m}_{sz}\\
	\dot{m}_{dx}\\
	\dot{m}_{dz}
	\end{matrix}\right) = \left(
	\begin{matrix}
	\alpha_s \dot{m}_{sz} + \tau_{sx} + \omega_{s2} m_{sz} - \alpha_s' \dot{m}_{dz}\\
	-\alpha_s \dot{m}_{sx} + \tau_{sz} - \omega_{s1} m_{sx} + \alpha_s' \dot{m}_{dx}\\
	\alpha_d \dot{m}_{dz} + \tau_{dx} + \omega_{d2} m_{dz} - \alpha_d' \dot{m}_{sz}\\
	-\alpha_d \dot{m}_{dx} + \tau_{dz} - \omega_{d1} m_{dx} + \alpha_d' \dot{m}_{sx}
	\end{matrix}\right).
\end{align}
To solve this, we first consider the system for the source and detector layers independently, treating the coupling terms between the source and detector layers as modifications to the nonequilibrium torques
\begin{align}
\left(
\begin{matrix}
\dot{m}_{sx}\\
\dot{m}_{sz}
\end{matrix}\right) &= \left(
\begin{matrix}
\alpha_s \dot{m}_{sz} + \omega_{s2} m_{sz} + \tau_{sx} - \alpha_s' \dot{m}_{dz}\\
-\alpha_s \dot{m}_{sx} - \omega_{s1} m_{sx} + \tau_{sz} + \alpha_s' \dot{m}_{dx}
\end{matrix}\right)\\
	\left(
\begin{matrix}
\dot{m}_{dx}\\
\dot{m}_{dz}
\end{matrix}\right) &= \left(
\begin{matrix}
\alpha_d \dot{m}_{dz} + \omega_{d2} m_{dz} + \tau_{dx} - \alpha_d' \dot{m}_{sz}\\
-\alpha_d \dot{m}_{dx} - \omega_{d1} m_{dx} + \tau_{dz} + \alpha_d' \dot{m}_{sx}
\end{matrix}\right)
\end{align}
and use the ansatz
\begin{equation}
	\bm_{i}(t) = \bm_{i} e^{-i\omega t}
\end{equation}
to arrive at the linearized form of the LLGS
\begin{align}
    \left(
    \begin{matrix}
    -i\omega m_{sx}\\
    -i\omega m_{sz}
    \end{matrix}\right) &= \left(
    \begin{matrix}
    -m_{sz}(i\omega \alpha_s - \omega_{s2}) + \tau_{sx} +i\omega\alpha_s' m_{dz}\\
    m_{sx}(i\omega\alpha_s - \omega_{s1}) + \tau_{sz} - i\omega\alpha_s' m_{dx}
    \end{matrix}\right)\\
    \left(
    \begin{matrix}
    -i\omega m_{dx}\\
    -i\omega m_{dz}
    \end{matrix}\right) &= \left(
    \begin{matrix}
    -m_{dz}(i\omega \alpha_d - \omega_{d2}) + \tau_{dx} +i\omega\alpha_d' m_{sz}\\
    m_{dx}(i\omega\alpha_d - \omega_{d1}) + \tau_{dz} - i\omega\alpha_d' m_{sx}
    \end{matrix}\right)
\end{align}
We can solve for the dynamics of each subsystem without decoupling as usual \cite{Karimeddiny2020}, which results in
\begin{align*}
	m_{sx} &= L_s \left(-\omega_{s2}\left[\tau_{sz} - i\omega \alpha_s' m_{dx}\right] + i\omega \left[\tau_{sx} + i\omega \alpha_s' m_{dz}\right]\right)\\
	m_{sz} &= L_s \left(\omega_{s1}\left[\tau_{sx} + i\omega \alpha_s' m_{dz}\right] + i\omega \left[\tau_{sz} + i\omega \alpha_s' m_{dx}\right]\right)\\
	m_{dx} &= L_d \left(-\omega_{d2}\left[\tau_{dz} - i\omega \alpha_d' m_{sx}\right] + i\omega \left[\tau_{dx} + i\omega \alpha_d' m_{sz}\right]\right)\\
	m_{dz} &= L_d \left(\omega_{d1}\left[\tau_{dx} + i\omega \alpha_d' m_{sz}\right] + i\omega \left[\tau_{dz} + i\omega \alpha_d' m_{sx}\right]\right)
\end{align*}
with $L_i = \left[\left(\omega^2-\omega_{i0}^2\right) + i\omega\omega_i^+\alpha_i\right]^{-1}$, $\omega_{i0}^2 = \omega_{i1}\omega_{i2}$ and $\omega_i^+ = \omega_{i1}+\omega_{i2}$.
We now need to decouple the source and detector layer resonances.
If we ignore any terms greater than first order in the damping parameters, this is relatively easy to do via substitution.
We will only solve for the detector layer magnetizations as this system is symmetric under interchanging $s\leftrightarrow d$.
The in-plane deflection becomes

\begin{equation}
    m_{dx} = L_d\left(-\omega_{d2}\tau_{dz} + i\omega\tau_{dx} + \alpha_d' L_s \left[ -\omega^2\tau_{sx}(\omega_{s1}+\omega_{d2})-i\omega\tau_{sz}(\omega_{s2}\omega_{d2}+\omega^2)\right]\right)\label{eq:mdx_first_sub}
\end{equation}
and the out-of-plane deflection becomes
\begin{equation}
    m_{dz} = L_d\left(-\omega_{d1}\tau_{dx} + i\omega\tau_{dz} + \alpha_d' L_s \left[\omega^2\tau_{sz}(\omega_{s2}-\omega_{d1})+i\omega\tau_{sx}(\omega_{d1}\omega_{s1} - \omega^2)\right]\right).
\end{equation}
We now separate the dynamic spin pumping term into real and imaginary parts. It will be convenient to define
\begin{equation}
	L_i = \frac{(\omega^2-\omega_{i0}^2)-i\omega\omega_i^+\alpha_i}{(\omega^2-\omega_{i0}^2)^2+\omega^2\omega_i^{+2}\alpha_i^2} \equiv \overline{L}_i\left((\omega^2-\omega_{i0}^2)-i\omega\omega_i^+\alpha_i\right)
\end{equation}
so that the in-plane deflections become
\begin{equation}\label{eq:mdx_reim}
    \begin{aligned}
	    m_{dx} = L_d &\left(-\omega_{d2} \tau_{dz} + \overline{L}_s\alpha_d'\left[(\omega_{s0}^2-\omega^2)\omega^2\tau_{sx}(\omega_{s1}+\omega_{d2}) - \omega^2\omega_s^+\alpha_s\tau_{sz}(\omega_{s2}\omega_{d2}+\omega^2)\right]\right.\\
	    &\ \left. +i\omega\left(\tau_{dx} + \overline{L}_s\alpha_d'\left[(\omega_{s0}^2-\omega^2)\tau_{sz}(\omega_{s2}\omega_{d2}+\omega^2) + \omega^2\omega_s^+\alpha_s\tau_{sx}(\omega_{s1}+\omega_{d2})\right]\right)\right)
	\end{aligned}
\end{equation}
and the out-of-plane deflections become
\begin{equation}\label{eq:mdz_reim}
    \begin{aligned}
        m_{dz} = L_d &\left(\omega_{d1}\tau_{dx} + \overline{L}_s\alpha_d'\left[(\omega_{s0}^2-\omega^2)\omega^2\tau_{sz}(\omega_{d1}-\omega_{s2}) + \omega^2\omega_s^+\alpha_s\tau_{sx}(\omega_{d1}\omega_{s1}-\omega^2)\right]\right.\\
        &\ \left.+i\omega \left(\tau_{dz} + \overline{L}_s \alpha_d'\left[(\omega_{s0}^2-\omega^2)\tau_{sx}(\omega^2-\omega_{d1}\omega_{s1}) +\omega^2 \omega_s^+\alpha_s\tau_{sz}(\omega_{d1}-\omega_{s2})\right]\right)\right).
    \end{aligned}
\end{equation}
Before converting to field units, we make two approximations. First, the latter part of the dynamic spin pumping contributions to Eq.~(\ref{eq:mdx_reim}) and Eq.~(\ref{eq:mdz_reim}) correspond to the symmetric part of the source layer Lorentzian.
This component falls off quickly away from the source resonant field and will be nearly zero near the detector layer resonance field, and are thus insignificant for our analysis.
Also for our system, the dampinglike torque on the source layer $\tau_{sx}$ is quite small, as is evidenced by the symmetric part of the source layer resonances being small compared with all other resonance components, including $\tau_{dz}$ in particular.
This justifies ignoring the occurrences of this torque appearing in the antisymmetric source layer Lorentzian in Eq.~(\ref{eq:mdx_reim}) and Eq.~(\ref{eq:mdz_reim}).
This leaves only one part of the dynamic spin pumping having a significant effect, and the in-plane and out-of-plane deflections can be written as
\begin{equation}\label{eq:mdx_reim_approx}
	m_{dx} = L_d \left(
	-\omega_{d2} \tau_{dz}
	+i\omega\left(\tau_{dx} + \overline{L}_s(\omega_{s0}^2-\omega^2)\alpha_d'\tau_{sz}(\omega_{s2}\omega_{d2}+\omega^2) \right)\right)
\end{equation}
and
\begin{equation}\label{eq:mdz_reim_approx}
m_{dz} = L_d \left(
\omega_{d1}\tau_{dx} + \overline{L}_s(\omega_{s0}^2-\omega^2)\alpha_d'\omega^2\tau_{sz}(\omega_{d1}-\omega_{s2}) 
+i\omega\tau_{dz}\right).
\end{equation}
We now wish to write the source resonance in field coordinates.
To do so we expand the resonant frequency about the source resonant field:
\begin{align}
    \omega_{s0}^2(B) &= \omega_{s0}^2\mid_{B=B_{s0}} + (B-B_{s0}) \left.\frac{d\omega_{s0}^2}{dB}\right|_{B=B_{s0}}\\
    &= \omega^2 + (B-B_{s0}) \left(\omega_{s1}\left.\frac{d\omega_{s2}}{dB}\right|_{B=B_{s0}}+\omega_{s2}\left.\frac{d\omega_{s1}}{dB}\right|_{B=B_{s0}}\right)\\
    &= \omega^2 + \gamma (B-B_{s0})\omega_s^+\mid_{B_{s0}}\\
    \omega_{s0}^2-\omega^2 &= \gamma (B-B_{s0})\omega_s^+\mid_{B_{s0}}
\end{align}
so in field coordinates
\begin{equation} \label{eq:field_lorentzian}
    \overline{L}_s(\omega_{s0}^2-\omega^2) = \frac{\gamma \omega_s^+\mid_{B_{s0}}(B-B_{s0})}{\gamma^2\omega_s^{+2}\mid_{B_{s0}}(B-B_{s0})^2+\omega^2\omega_s^{+2}\alpha_s^2} = \frac{1}{\alpha_s \omega_s^+\mid_{B_{s0}}}\frac{1}{\omega} \frac{(B-B_{s0})\Delta_s}{(B-B_{s0})^2+\Delta_s^2} \equiv \overline{L}_{sA}
\end{equation}
where we have defined $\Delta_i \equiv \omega\alpha_i/\gamma$ which is the linewidth of the resonance associated with layer $i$.

We can extract the symmetric and antisymmetric detector-layer resonance amplitudes by taking the real part of Eq.~(\ref{eq:mdx_reim_approx}) and Eq.~(\ref{eq:mdz_reim_approx}) combined with Eq.~(\ref{eq:field_lorentzian}) and identifying the prefactors on the symmetric and antisymmetric Lorentzian shapes, as done in more detail in \cite{Karimeddiny2020}.
Doing so, we find that
\begin{align}
    A_{dx} &= \frac{1}{\alpha_d\omega_d^+}\frac{1}{\omega}\omega_{d2}\tau_{dz}\label{eq:Adx_simple}\\
    S_{dx} &= \frac{1}{\alpha_d\omega_d^+}\left(\tau_{dx} + \overline{L}_{sA}(\omega_{s2}\omega_{d2}+\omega^2)\tau_{sz}\alpha_d'\right)\label{eq:Sdx_simple}\\
    A_{dz} &= \frac{-1}{\alpha_d\omega_d^+}\frac{1}{\omega}\left(\omega_{d1}\tau_{dx} + \overline{L}_{sA}\omega^2(\omega_{d1}-\omega_{s2})\tau_{sz}\alpha_d'\right)\label{eq:Adz_simple}\\
    S_{dz} &= \frac{1}{\alpha_d\omega_d^+}\tau_{dz} \label{eq:Sdz_simple}
\end{align}
with $S$ and $A$ representing the symmetric and antisymmetric amplitudes and the subscript $x$ and $z$ representing that those quantities deal with the in-plane and out-of-plane magnetization deflection amplitudes, respectively. 
Everything with a field dependence is evaluated at $B_{d0}$, except where noted in $\overline{L}_{sA}$, since all of these resonance amplitudes are measured close to the detector layer resonance field.
The effect of dynamic spin pumping captured in Eqs.~(\ref{eq:Sdx_simple}) and (\ref{eq:Adz_simple}) comes from spins pumped by the source layer which travel to and exert torques on the detector layer.
In the next section, we deal with another slightly more subtle effect which the dynamic spin pumping has on our measurements.

\subsection{Dynamic Spin Pumping from Detector Layer to Source Layer}\label{ssec:dynamSP_DS}

Here we show spins pumped from the detector layer to the source layer also end up having an effect on our measurement of the detector-layer resonance lineshape.
To see that, we will start at the equivalent of Eq.~(\ref{eq:mdx_first_sub}) for the source layer magnetization (simply replacing $d\leftrightarrow s$) and rearrange some terms
\begin{align}
    m_{sx} &= L_s\left(-\omega_{s2}\tau_{sz} + i\omega\tau_{sx} + \alpha_s' L_d \left[
    -\omega^2\tau_{dx}(\omega_{d1}+\omega_{s2})-i\omega\tau_{dz}(\omega_{d2}\omega_{s2}+\omega^2)\right]\right)\\
    &= L_s(-\omega_{s2}\tau_{sz} + i\omega \tau_{sx}) + \alpha_s'L_dL_s\left[
    -\omega^2\tau_{dx}(\omega_{d1}+\omega_{s2})-i\omega\tau_{dz}(\omega_{d2}\omega_{s2}+\omega^2)\right]\\
    &= L_s(-\omega_{s2}\tau_{sz} + i\omega \tau_{sx}) + \alpha_s'L_d \overline{L}_s\begin{aligned}
    &\left[(\omega_{s0}^2-\omega^2)\omega^2(\omega_{d1}+\omega_{s2})\tau_{dx} - \omega^2\omega_s^+(\omega_{d2}\omega_{s2}+\omega^2)\alpha_s\tau_{dz}\right.\\
    &\ \left.+i\omega\left((\omega_{s0}^2-\omega^2)(\omega_{d2}\omega_{s2}+\omega^2)\tau_{dz}+\omega^2\omega_s^+(\omega_{d1}+\omega_{s2})\alpha_s\tau_{dx}\right)\right].\label{eq:msx_real_full}
    \end{aligned}
\end{align}
The first part of Eq.~(\ref{eq:msx_real_full}) is the usual equation for the source layer resonance, with the second part being due to the dynamic spin pumping.
Near the detector resonance, the first term in Eq.~(\ref{eq:msx_real_full}) is essentially a constant offset, but the $L_d$ term means that the source layer magnetization will resonate at the deteciton layer's resonance field and with the detector layer's linewidth.
The voltage we measure is still due to rectification from the soure layer AMR mixing with the applied current, but since we measure the voltage across the whole heterostructure this signal will add to the resonance we measure at the detector-layer's resonant field and linewidth.

To see how this affects things, we can make a few simplifications again. If we want to look near the detector layer resonance, the terms related to the symmetric part of the source layer resonance (on the right in the square brackets in Eq.~(\ref{eq:msx_real_full}) will be negligible (same argument as above), so we can ignore them, letting the source layer oscillation near the detector layer resonance look like 
\begin{equation}\label{eq:msx_simplified}
    m_{sx} = L_s(-\omega_{s2}\tau_{sz} + i\omega \tau_{sx}) + L_d\left[\overline{L}_{sA}\omega^2(\omega_{d1}+\omega_{s2})\tau_{dx}\alpha_s' + i\omega\overline{L}_{sA} (\omega_{d2}\omega_{s2}+\omega^2)\tau_{dz}\alpha_s'\right].
\end{equation}
Since we are only concerned with this signal near the detector-layer resonance, and the first term in Eq.~(\ref{eq:msx_simplified}) is essentially a constant offset, we ignore it in the following analysis.
We can decompose the signal from the second part into symmetric and antisymmetric parts as above. 
We find that
\begin{align}
    \Tilde{A}_d &= -\frac{1}{\alpha_d\omega_d^+}\frac{1}{\omega}\overline{L}_{sA}\omega^2(\omega_{d1}+\omega_{s2})\tau_{dx}\alpha_s' \label{eq:Ads_simple}\\
    \Tilde{S}_d &= \frac{1}{\alpha_d\omega_d^+}\overline{L}_{sA}(\omega_{d2}\omega_{s2}+\omega^2)\tau_{dz}\alpha_s'\label{eq:Sds_simple}
\end{align}
where the $\sim$ signifies that this appears as additions to the detector-layer resonance since these lorentzians have the same resonant field and lineshape, but actually come from the source layer.

\subsection{Overall Effect on Measured Detector-Layer Resonance}\label{ssec:dynamSP_full}

Combining Eqs.~(\ref{eq:Adx_simple}) and(\ref{eq:Ads_simple}), and Eqs.~(\ref{eq:Sdx_simple}) and (\ref{eq:Sds_simple}), keeping note of which layer gives rise to the rectification producing the voltage we measure, we find the expression for the voltage we measure when fitting the resonance which has the resonant field and linewidth of the detector layer:
\begin{align}
    V_{dS} &= \frac{IR_d^\text{AMR}\sin(2\phi)}{2}\frac{1}{\alpha_d\omega_d^+}\left[\tau_{dx}+\overline{L}_{sA}(\omega_{s2}\omega_{d2}+\omega^2)\tau_{sz}\alpha_d'\right] \\
    &+ \frac{IR_s^\text{AMR}\sin(2\phi)}{2}\frac{1}{\alpha_d\omega_d^+}\overline{L}_{sA}(\omega_{s2}\omega_{d2}+\omega^2)\tau_{dz}\alpha_s'\label{eq:VdS_full2}\\
    V_{dA} &= \frac{IR_d^\text{AMR}\sin(2\phi)}{2} \frac{1}{\alpha_d\omega_d^+}\frac{1}{\omega}\omega_{d2}\tau_{dz} - \frac{IR_s^\text{AMR}\sin(2\phi)}{2}\frac{1}{\alpha_d\omega_d^+}\frac{1}{\omega}\alpha_s'\overline{L}_{sA}\omega^2(\omega_{d1}+\omega_{s2})\tau_{dx}\label{eq:VdA_full2}.
\end{align}
with $I$ the total current in the heterostructure and and $R_i^\text{AMR}$ the resistance change of the whole stack due to the AMR of layer $i$.

In the conventional ST-FMR setting with only one ferromagnetic layer, we could look at the ratio of $V_{dS}$ and $V_{dA}$ to find the charge-to-spin conversion efficiency, but with the added terms this ratio is now harder to interpret, so we will work to simplify this expression a bit and see what information we can get from this ratio.
We first try to simplify our expression for $V_{dA}$.
We can write it as 
\begin{equation}
    V_{dA} = \frac{I_dR_d^\text{AMR}\sin(2\phi)}{2}\frac{1}{\alpha_d\omega_d^+}\frac{1}{\omega}\omega_{d2}\tau_{dz}\left(1-\frac{ R_s^\text{AMR}}{R_d^\text{AMR}} \frac{\overline{L}_{sA}\omega^2 (\omega_{d1}+\omega_{s2})}{\omega_{d2}}\frac{\tau_{dx}}{\tau_{dz}}\alpha_s'\right)
\end{equation}
and will now show that the term subtracting from 1 is small.
For the Co/Py system, all of the terms subtracting 1 are known or can be reasonably approximated.
First, $\overline{L}_{sA}\omega^2 (\omega_{d1}+\omega_{s2})/\omega_{d2}\approx -1$ which we can find from fitting the resonances.
We estimate the torque ratio to be in the range $0.2-0.5$. 
The AMR ratio is close to 1 from our AMR measurements in Section \ref{sec:AMR}.
The spin pumping coupling term $\alpha_s'$ should be on the same order as damping in Py, so estimating $0.01$ should be reasonable.
Putting all of this together, the whole term to the right of the 1 inside the parentheses comes out to no more than $0.01$, so this should be no more than a 1\% correction to $V_{dA}$.
Therefore, we ignore the effect of dynamic spin pumping on $V_{dA}$, which will simplify our analysis.

After simplifying $V_{dA}$, the $V_{dS}/V_{dA}$ ratio becomes
\begin{equation}
    \frac{V_{dS}}{V_{dA}} = \frac{\omega}{\omega_{d2}}\left[\frac{\tau_{dx}}{\tau_{dz}}+\overline{L}_{sA}(\omega_{s2}\omega_{d2}+\omega^2)\frac{\tau_{sz}}{\tau_{dz}}\left(\alpha_d' + \frac{\tau_{dz}}{\tau_{sz}}\frac{R^\text{AMR}_s}{R^\text{AMR}_d}\alpha_s'\right)\right]
\end{equation}
and we can then define
\begin{equation}\label{eq:xiFMR_full}
    \xiFMR = \xiDL + \eta \overline{L}_{sA}(\omega_{s2}\omega_{d2}+\omega^2)\frac{\tau_{sz}}{\tau_{dz}}\left(\alpha_d' + \frac{\tau_{dz}}{\tau_{sz}}\frac{R^\text{AMR}_s}{R^\text{AMR}_d}\alpha_s'\right)
\end{equation}
which is the expression that appears in the main text.

As discussed in the main text, the torque ratio can be expressed as $\tau_{sz}/\tau_{dz}=-(1-x_s)/x_s$ where the shunt factor $x_s$ is the fraction of the total RF current flowing within the source layer. Based on a parallel-conduction model using the layer resistances determined as described above, the value of $x_s$ as a function of Py thickness is shown in Supplementary Fig.\ \ref{fig:S4}.

\begin{figure*}
\begin{center}
\includegraphics[width=0.6\linewidth]{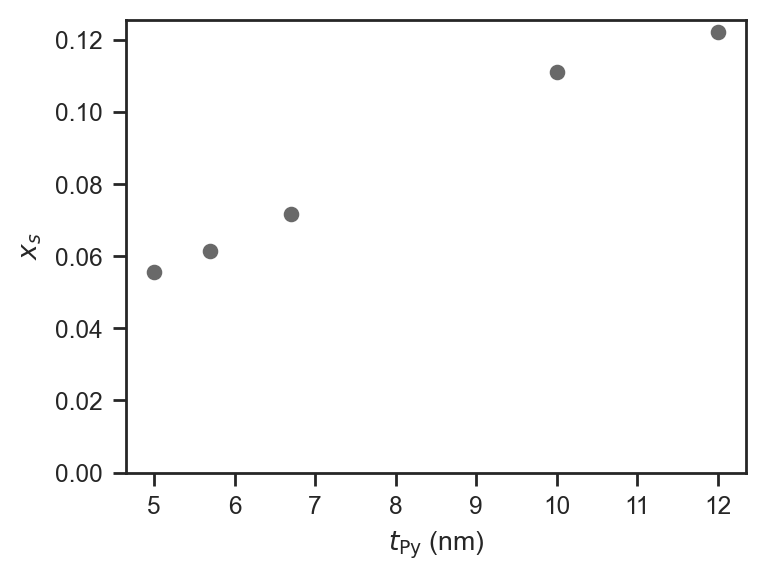}
\caption{Source layer thickness dependence of the fraction of current flowing in the Py layer, $x_s$.
}
\label{fig:S4}
\end{center}
\end{figure*}

\end{document}